\documentclass[%
 reprint,
 amsmath,amssymb,
 aps,
]{revtex4-2}

\usepackage{graphicx}
\usepackage{dcolumn}
\usepackage{bm}
\usepackage{xcolor}
\usepackage{soul}

\begin{document}

\preprint{APS/123-QED}

\title{Strongly nonlinear nanocavity exciton-polaritons in gate-tunable monolayer semiconductors} 

\author{Zhi Wang}
\thanks{These authors contributed equally}%
\author{Li He$^*$}%
\email{heli1@sas.upenn.edu}
\author{Bumho Kim}%
\author{Bo Zhen}%
\email{bozhen@sas.upenn.edu} 
\affiliation{%
 Department of Physics and Astronomy, University of Pennsylvania, Philadelphia, PA 19104, USA
}%




\date{\today}

\begin{abstract}
Strong coupling between light and matter in an optical cavity provides a pathway to giant polariton nonlinearity, where effective polariton-polariton interactions are mediated by materials' nonlinear responses. 
The pursuit of such enhanced nonlinearity at low optical excitations, potentially down to the single-particle level, has been a central focus in the field, inspiring the exploration of novel solid-state light-matter systems. 
Here, we experimentally realize extremely nonlinear and robust cavity exciton-polaritons by coupling a charge-tunable MoSe$_2$ monolayer to a photonic crystal nanocavity.
We show that the observed polariton nonlinearity arises from increased exciton dephasing at high populations, leading to diminished exciton-photon coupling and ultimately the breakdown of the strong coupling condition.
Remarkably, the strong mode confinement of the nanocavity enables all-optical switching of the cavity spectrum at ultralow optical excitation energies, down to \textcolor{black}{$\sim 4$} fJ, on picosecond timescales.
Our work paves the way for further exploration of 2D nonlinear exciton-polaritons, with promising applications in both classical and quantum all-optical information processing.

\end{abstract}

\maketitle



The coherent coupling between excitonic resonances in solid-state materials and optical fields within an optical cavity gives rise to quasiparticles known as exciton-polaritons, which exhibit dual characteristics of light and matter. 
These hybrid light-matter particles underpin a wide range of fascinating  classical and quantum phenomena 
, such as polariton lasing \cite{schneider2013electrically}, polariton Bose-Einstein condensation\cite{byrnes2014exciton}, and superfluidity \cite{amo2009superfluidity}. 
Among their many remarkable attributes, a particularly appealing aspect of exciton-polaritons is their unprecedented nonlinearity, originating from the excitonic constituents.
This allows the observation of nonlinear optical phenomena at extremely low light intensities, even down to the level of single photons \cite{hennessy2007quantum,englund2007controlling}. 
Harnessing the substantial polariton nonlinearity in solid-state systems could establish a robust and scalable pathway for on-chip all-optical computing and photonic quantum information processing.

Recent advancements in two-dimensional (2D) transition metal dichalcogenides (TMDs) \cite{wang2018colloquium} and their integration with photonic structures \cite{mak2016photonics}
 have provided new opportunities for exploring nonlinear exciton-polaritons at the atomic scale. 
In the study of 2D exciton-polariton systems, achieving tightly confined optical modes in all three dimensions is crucial for enhancing both light-matter and matter-matter interactions. 
For example, when interfacing a uniform 2D TMD flake with a planar optical cavity (assuming the flake size is comparable to the lateral dimensions of the cavity mode), the vertical optical confinement plays a key role in strengthening exciton-photon interactions and the resulting Rabi splitting ($\Omega \sim d^{-1/2}$, where $d$ is the effective length of the optical mode in the vertical direction). 
On the other hand, effective polariton-polariton interactions can arise at high polariton populations either from short-range exciton-exciton interactions \cite{delteil2019towards,munoz2019emergence} or from depletion of excitonic states (also known as the phase space filling effect)  \cite{emmanuele2020highly,zhang2021van}.
Confining optical modes to a compact lateral area increases the spatial overlap of excitons and reduces the available exciton states in the cavity, thereby significantly amplifying polariton nonlinearity.
Recently, nonlinear exciton-polaritons have been extensively studied in various 2D material systems \cite{tan2020interacting,emmanuele2020highly,zhang2021van,gu2021enhanced,datta2022highly}.
However, most research has relied on bulky optical cavities, such as distributed Bragg reflectors, where the polariton modes are extended in one or more dimensions, resulting in diminished exciton-photon and polariton-polariton interactions. 
The insufficient mode confinement remains a significant barrier to further lowering the polariton nonlinear threshold, placing quantum polariton nonlinearity far beyond current experimental reach.

Here we demonstrate robust and highly nonlinear 2D exciton-polaritons by integrating a gate-tunable MoSe$_2$  monolayer with a silicon nitride photonic crystal (PhC) nanocavity. 
Notably, our nanocavity boasts an extremely compact mode volume \textcolor{black}{$\sim 1.3(\lambda /n)^{3}$}, orders of magnitude smaller than those in prior studies of 2D exciton-polaritons. 
The enhanced mode confinement in all three dimensions boosts the exciton-photon coupling strength, resulting in a large Rabi splitting, while significantly lowering the polariton nonlinear threshold.
At low optical excitation, the coupled TMD-PhC nanocavity facilitates strong hybridization between excitons and cavity photons, forming cavity exciton-polaritons. 
With increasing optical excitation, we observe a significant rise in the excitons dephasing rate due to their incoherent coupling with an exciton reservoir \cite{takemura2015dephasing,takemura2016coherent,moody2015intrinsic}. This excitation-induced dephasing (EID) effectively reduces Rabi splitting, resulting in nonlinear shifts in polariton energies and eventually the collapse of the strong coupling condition.
\textcolor{black}{Remarkably}, we achieve all-optical switching of the cavity spectrum at an energy level as low as $\sim 4$ fJ, setting a new benchmark for switching energy in 2D exciton-polariton systems.
Through pump-probe measurements, we further confirm that the nonlinear polariton dynamics occur on picoseconds timescales. 
Our work represents a significant step towards the realization of robust and scalable nonlinear polaritonics on an integrated platform for both classical and quantum applications.

%

\begin{figure}[t]
    \centering
    \includegraphics[width=1\columnwidth]{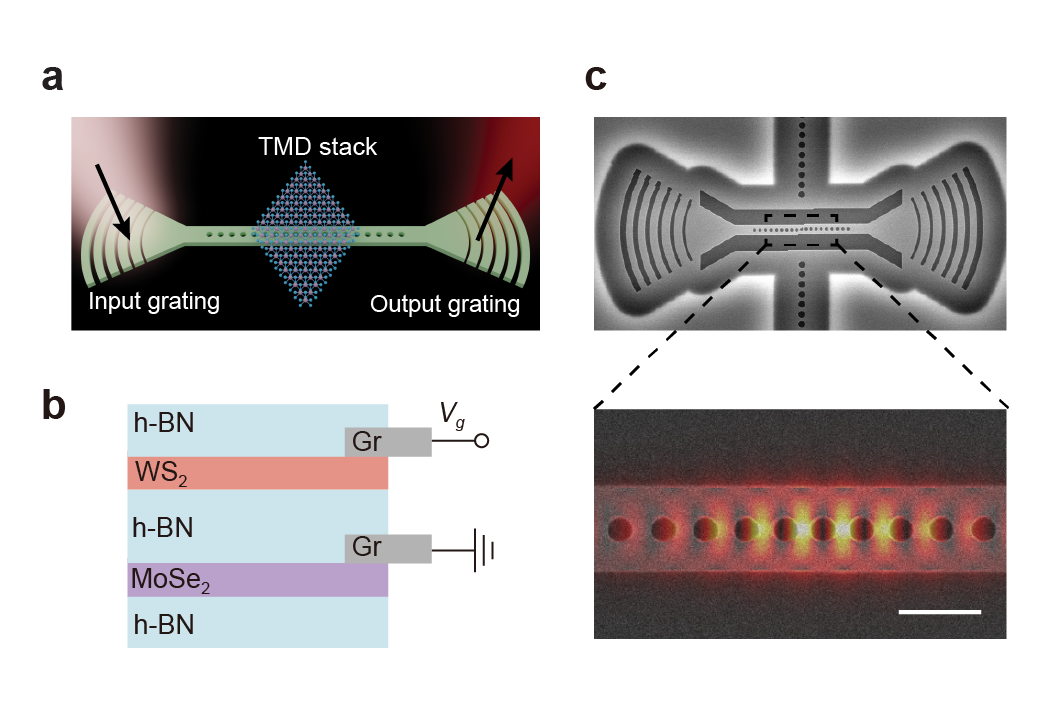} 
    \caption{\label{fig:epsart} 2D nanocavity exciton-polaritons. a. Schematic of the coupled TMD-PhC nanocavity. b. Schematic of the gate-tunable TMD stack. c. Scanning electron microscope image of the suspended Si$_3$N$_4$ nanobeam cavity, with the inset showing the simulated cavity mode profile. Scale bar, 500 nm.}
\end{figure}

\begin{figure*}[t]
    \centering
    \includegraphics[width=2\columnwidth]{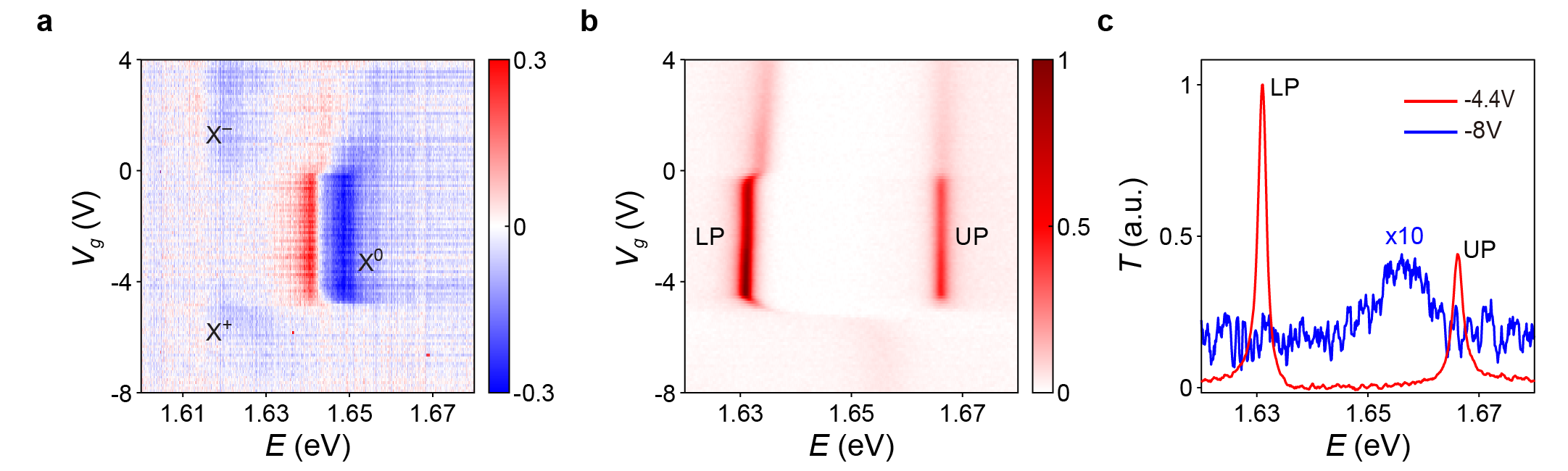}    \caption{\label{fig:wide} Linear characterization of the coupled TMD-PhC nanocavity. a, Gate-dependent reflectance spectrum ($dR/R$) features excitons in the charge-neutral regime and trions in the doped  regime. The plot is normalized to the reflectance spectrum at $V_g = -10 V$, where no excitonic features are observed. b, Gate-dependent cavity transmission spectrum measured through grating couplers. c, Transmission spectra measured in the charge-neutral (red) and p-doped regime (blue).}
\end{figure*}

Our TMD-PhC nanocavity is schematically shown in Fig. 1a. The TMD stack consists of MoSe$_2$ and WS$_2$ monolayers separated by a $\sim$10 nm thick hBN (Fig. 1b). Our work focuses on intralayer excitons in the MoSe$_2$ monolayer, with the WS$_2$ monolayer serving as a transparent top gate that avoids significant optical absorption near the MoSe$_2$ exciton energy. Both TMD monolayers are contacted with graphite electrodes for electrostatic gating.
To achieve strong exciton-photon coupling, we fabricate a photonic crystal nanobeam cavity on a silicon nitride-on-insulator wafer with a 145 nm thick silicon nitride layer.
Figure 1c shows the scanning electron microscope image of the nanobeam cavity.
The whole device is suspended from the substrate to increase refractive index contrast.
Light is coupled into and out of the nanobeam cavity using a pair of grating couplers at the two ends.
In the absence of the TMD stack, the bare cavity exhibits an optical resonance at \textcolor{black}{1.736} eV with a full-wdith-half-maximum linewidth of \textcolor{black}{$\gamma_{\text{cav}} = $ 1.9 meV}.
The cavity linewidth is determined by its radiative coupling to the waveguides on both sides ($\gamma_\text{rad} \sim 0.7$ meV) and optical loss due to scattering from fabrication disorders and material absorption ($\gamma_\text{nonrad} \sim 1.2$ meV). 
The TMD stack is then transferred onto the photonic chip with pre-patterned gold electrodes using standard dry transfer technique. 
The presence of the TMD stack is expected to red-shift the cavity resonance
, resulting in a cavity-exciton detuning \textcolor{black}{$\delta_\text{c-X} \sim 10$} meV at cryogenic temperatures.
Meanwhile, it is also expected to increase the cavity-waveguide coupling to \textcolor{black}{$\gamma_{\text{rad}} \sim$ 1.4 meV}, owing to the reduced photonic band gap confinement (see SI for more information). 
Our design yields a cavity-waveguide coupling strength comparable to the intrinsic MoSe$_2$ exciton linewidth of $\gamma _X \sim$1 meV \cite{kim2021free,zhou2020controlling}. This ensures narrow polariton linewidths when excitons are strongly coupled to the cavity resonance, while simultaneously enabling robust optical readout signal through the coupling waveguides.

\begin{figure*}
    \centering
     \includegraphics[width=2\columnwidth]{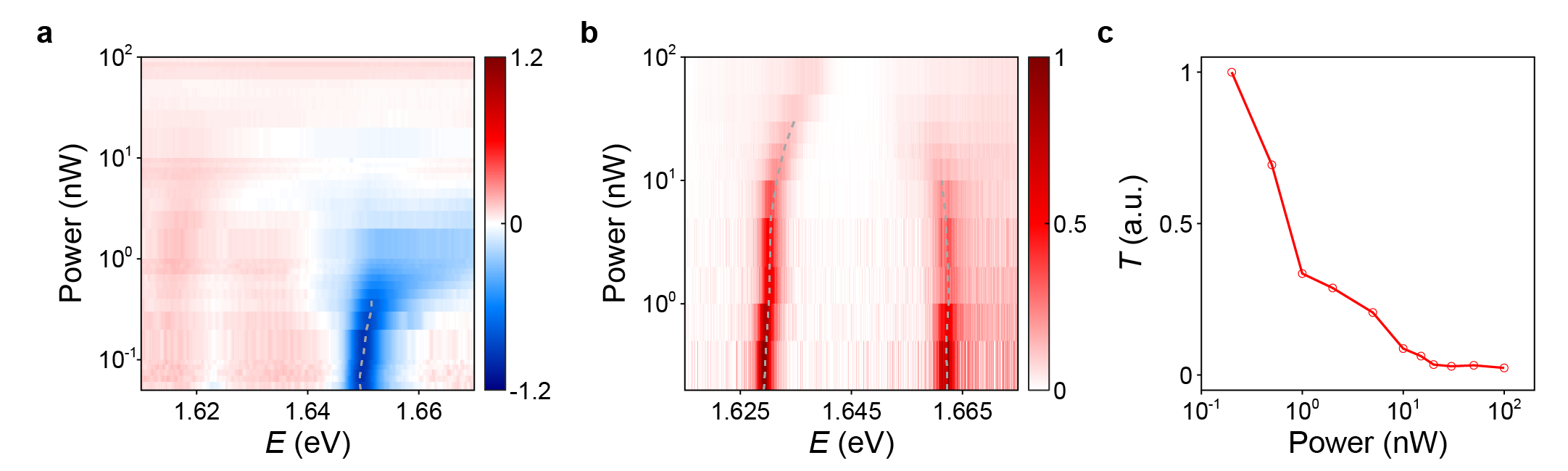}  
    \caption{\label{fig:wide} Nonlinear cavity exciton-polaritons. a. Exciton reflectance and b. cavity transmission measured as a function of excitation power. The gray dashed lines indicate the evolution of exciton and polariton energies. c. Plot of cavity transmission at the equilibrium LP energy ($E_\text{LP}=1.629$ eV) versus optical excitation.}
    \end{figure*}

The hybrid device is measured at 4K. We first characterize the linear optical properties of MoSe$_2$ excitons through gate-dependent reflectance measurement in a confocal microscope setup: the MoSe$_2$ flake is illuminated with a broadband light source from the normal direction and the reflected light is collected. The reflectance spectrum shown in \textcolor{black}{Fig. 2a} reveals complex excitonic features that vary significantly with the doping level in the MoSe$_2$ monolayer \cite{ross2013electrical,li2021refractive}. Specifically, in the charge-neutral regime \textcolor{black}{ ($-4.5V < V_g < 0V$),} the optical properties of MoSe$_2$ are dominated by a strong neutral exciton resonance ($X^0$) at \textcolor{black}{1.646} eV. 
Increasing or decreasing $V_g$ to above $0 V$ or below $-3 V$ brings the MoSe$_2$ monolayer into n-doped or p-doped regimes, respectively, where the exciton resonance vanishes and weak trion resonances ($X^\pm$) emerge at \textcolor{black}{1.616} eV.

Next, we study how neutral excitons hybridize with nanocavity resonance to form cavity exciton-polaritons through transmission measurements. To this end, light is injected into the waveguide and the transmitted light is collected off-chip through grating couplers. The measured gate-dependent transmission spectrum is shown in \textcolor{black}{Fig. 2b}. 
In the charge-neutral regime, where the exciton oscillator strength is maximal, the cavity transmission spectrum features two prominent peaks at \textcolor{black}{1.666 eV and 1.631 eV}, corresponding to the upper (UP) and lower (LP) polariton states, respectively. 
The measured linewidths are \textcolor{black}{2.3} meV for the UP state and \textcolor{black}{1.8} meV for the LP state. 
In contrast, as the system is gradually tuned into the p-doped regime ($V_g < -4.5V$), the two polariton peaks merge into a single peak around \textcolor{black}{1.656} eV, corresponding to a significantly damped cavity resonance. 
This strong to weak coupling transition indicates weakening of the exciton-photon coupling strength with increasing doping level, consistent with the previous reflection measurement (Fig. 2a). 
The measured broad cavity linewidth of \textcolor{black}{9.9} meV, along with the suppressed optical transmission in the p-doped regime, suggests increased optical absorption at high doping levels (Fig. 2c).
A similar trend of strong to weak coupling transition is also observed when electron doping is introduced into the MoSe$_2$ monolayer ($V_g > 0V$).
Based on the coupled oscillator model, we estimate the exciton-photon coupling strength in the charge-neutral regime is \textcolor{black}{$g = 16.8$} meV, reflecting the strong vertical field confinement of the nanocavity due to index confinement. 

\begin{figure}[t]
    \centering
    \includegraphics[width=1\columnwidth]{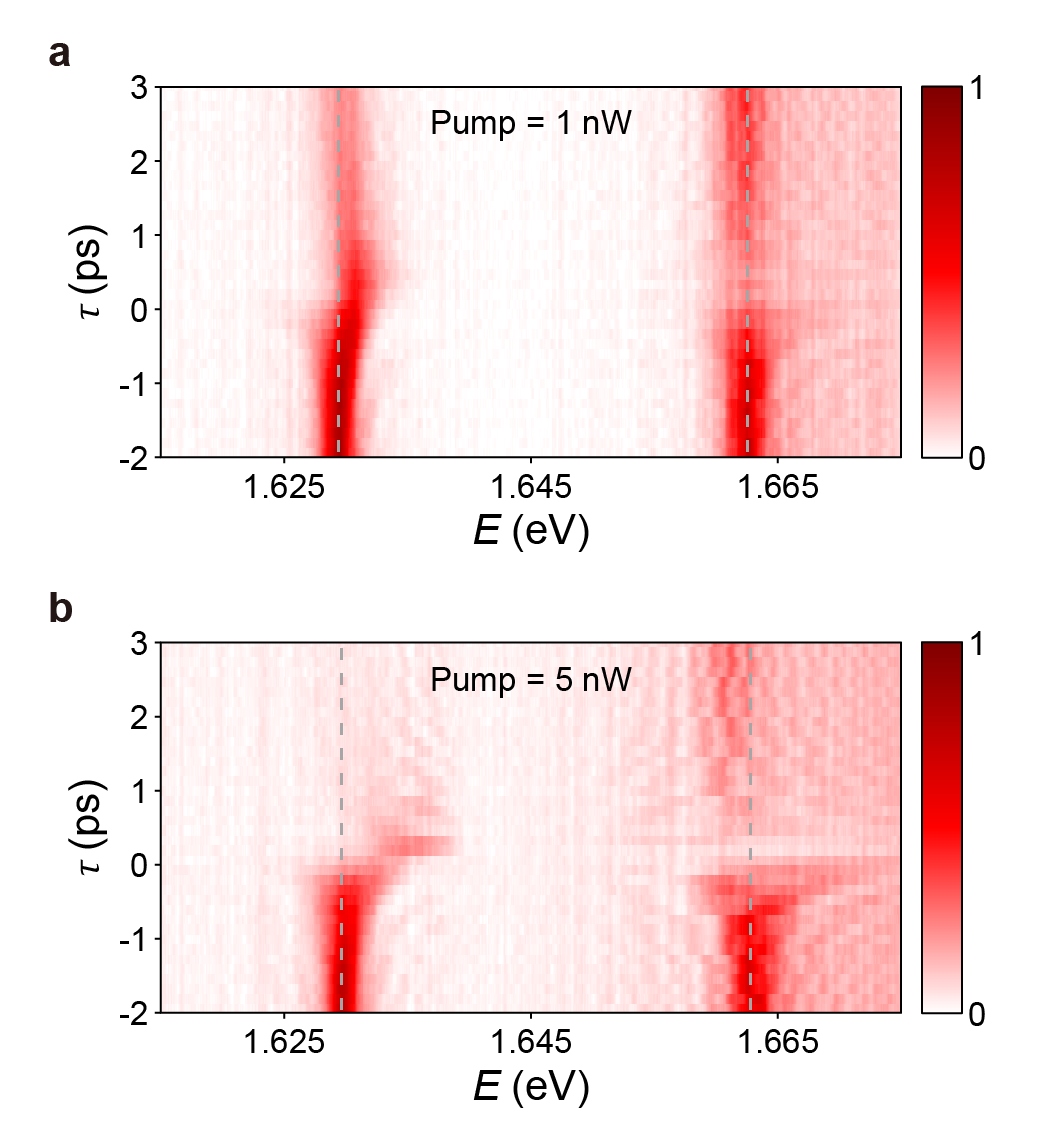} 
    \caption{\label{fig:epsart} Pump-probe spectroscopy of nonlinear cavity polaritons at two distinct excitations. The gray dashed lines represent the equilibrium energies of the LP and UP states.}
\end{figure}


We then investigate the nonlinear responses of excitons and cavity polaritons under excitation with femtosecond laser pulses (pulse duration $\sim$50 fs, 3 kHz repetition rate). 
To characterize the nonlinear behavior of MoSe$_2$ excitons, we tightly focus the laser beam on the sample and measure reflectance spectrum as a function of optical power.
As shown in Fig. 3a, in the charge-neutral regime
, a sharp exciton resonance is observed for optical excitation levels below \textcolor{black}{0.1} nW, consistent with the previous linear characterization. 
However, as the optical power increases from \textcolor{black}{0.1} nW to 0.5 nW, the exciton linewidth starts to broaden, accompanied by a 2 meV blue shift in exciton energy, while the exciton oscillator strength remains unchanged (see SI for more information). 
Further increasing the excitation above 0.5 nW results in the gradual diminishing of the exciton feature. 
The observed blue shift in exciton resonance stems from energy renormalization driven by increased exciton-exciton interactions at high exciton densities.
On the other hand, the linewidth broadening is attributed to excitation-induced dephasing, where the coherence of optically excited excitons is disrupted due to their increasing coupling with the incoherent exciton reservoir through exciton-exciton interactions.
Such EID effect has been widely observed in conventional 3D and quasi-2D semiconductors \cite{takemura2015dephasing,takemura2016coherent}, as well as in similar 2D TMD monolayers \cite{moody2015intrinsic}. 


To explore the impact of the observed exciton nonlinearity on cavity responses, we measure how the cavity transmission spectrum varies with optical power. 
As shown in \textcolor{black}{Fig. 3b}, under low optical excitation, the LP and UP peaks are slightly red-shifted by a few meV relative to the linear measurement in Fig. 2b, due to the slow temporal drift of the cavity resonance (see SI for more information).
Increasing the optical power from \textcolor{black}{0.2 nW to 20 nW} (measured before the input grating coupler) results in a blue shift in the LP energy, while the energy shift in UP state is less pronounced. 
Moreover, the linewidths of both polariton modes broaden as excitation power increases, ultimately leading to the breakdown of the strong-coupling condition above \textcolor{black}{100} nW. 
In this regime, the coupled TMD-PhC nanocavity behaves as a damped optical resonator, effectively suppressing optical transmission. 
We note that the cavity polaritons show a less pronounced power dependence compared to excitons. This is because the increase in polariton linewidths reduces the waveguide-cavity coupling effiency, leading to sub-linear dependence of the polariton density on excitation power. In contrast, in the free-space experiment (Fig. 3a), the exciton density scales linearly with optical power.

The nonlinear cavity response is a direct consequence of increased exciton dephasing at high excitation densities.
This can be qualitatively understood using the coupled oscillator model, where the system's two complex eigenvalues are given by
\begin{eqnarray*}
E_{\pm} = (E_X + E_\text{cav})/2 - i (\gamma_X +\gamma_\text{cav})/4 \\  \nonumber
\pm \sqrt{g^2  + \frac{1}{4} \Bigr[ \delta_\text{c-X} - i \frac{(\gamma_\text{cav} - \gamma_X)}{2} \Bigr]^2}.
\end{eqnarray*} 
Increasing polariton density leads to a rise in both exciton energy $E_X$ and dephasing rate $\gamma_X$.
The former causes a blue shift in both LP and UP states, while the latter leads to a blue shift in the LP and a red shift in the UP state due to reduced Rabi splitting $\Omega = \operatorname{Re} \Bigl\{ 2 \sqrt{g^2  + \frac{1}{4} [\delta_\text{c-X} - i \frac{(\gamma_\text{cav} - \gamma_X)}{2} ]^2} \Bigr\}$.
This explains the asymmetric energy shifts observed for the LP and UP states as the optical excitation increases.
Furthermore, when the exciton dephasing rate approaches the exciton-photon coupling strength \textcolor{black}{$g$}, the strong coupling condition collapses, and the cavity spectrum becomes dominated by a damped optical resonance.
A more rigorous theoretical analysis of the nonlinear polariton properties requires including the incoherent couplings between the two polariton states and the exciton reservoir, extending beyond the minimal coupled oscillator model \cite{takemura2015dephasing,takemura2016coherent}.
Notably, the spectral width of the laser pulses (\textcolor{black}{$\sim 60$ meV}) is significantly broader than the polariton linewidths. As a result, only a small fraction of the pulse energy is effectively coupled into the cavity to generate the polariton population. 
This suggests that the realistic switching energy --- the pulse energy required to shift the polariton resonance by half its linewidth --- can be reduced by at least an order of magnitude, reaching as low as \textcolor{black}{$\sim 4$} fJ when the spectral width of the laser pulses is matched to the polariton linewidths (see SI for more information). 

Finally, we demonstrate that the nonlinear polariton dynamics occur at ultrafast time scales through a pump-probe measurement. 
In this setup, a pump beam is focused normally onto the cavity center to generate an exciton population, while a probe beam, derived from the same laser source, is coupled through the grating coupler to measure cavity transmission spectrum as a function of time delay. 
Note that the pump beam does not directly couple to the cavity polaritons, as these modes exhibit negligible radiative coupling with free space.
Figure 4 shows the cavity transmission spectra as a function of time delay at two different excitation powers. 
At low excitation of 1 nW (Fig. 4a), the LP state exhibits a maximal blue shift \textcolor{black}{of 1.3 meV} at positive delay, after which it gradually returns to its equilibrium energy in $\sim$ 1ps, consistent with the measured polariton linewidth of \textcolor{black}{1.8} meV. 
In comparison, the UP mode displays no significant energy shift, similar to the previous power-dependent cavity spectrum measurement (Fig. 3b).
We also observe a decrease in both LP and UP transmissions at positive delay, with full recovery to equilibrium states taking on the order of hundreds of picoseconds (see SI for additional information). 
This slow recovery is likely attributed to the excitation of impurity-trapped exciton states, which have lower energy and relatively long lifetime on the order of 100 ps \cite{kim2021free}.
However, further studies are needed to better understand the origin of these long-lived excitations and to investigate potential strategies, such as utilizing high-quality bulk crystals with low defect densities, to mitigate the prolonged recovery time.
In contrast, at high excitation level of \textcolor{black}{5 nW} (Fig. 4b), the cavity polariton modes exhibit dramatic energy shifts, followed by the quenching of the cavity transmission within $\sim 200$ fs, indicating the breakdown of the strong coupling condition. 

By tightly confining 2D exciton-polaritons to the subwavelength scale using PhC nanocavities, we achieved extremely nonlinear cavity polaritons. This enabled all-optical switching of cavity spectrum at a record-low pulse energy of 4 fJ, corresponding to a photon number of $N\sim 10^4$. 
We anticipate that this nonlinear threshold can be further reduced by several orders of magnitude, potentially reaching the regime of quantum nonlinearity, through the use of 2D TMD materials with enhanced excitonic nonlinearity and optimized optical cavity designs.
For instance, 2D systems with reduced excitonic density of states, such as trions in lightly doped TMD monolayers \cite{tan2020interacting,emmanuele2020highly} or intralayer moiré excitons in TMD heterobilayers \cite{zhang2021van}, can exhibit substantially enhanced exciton nonlinearity due to phase-space filling --- the saturation of excitonic states under high optical excitations. Incorporating these charge and interlayer coupling degrees of freedom into our hybrid TMD-PhC platform is expected to reduce the nonlinear threshold by up to two orders of magnitude. 
Additionally, the optical confinement in our nanobeam cavity is limited by the relatively low refractive index of silicon nitride ($n=2.15$) and material absorption.
Transitioning to high-index-contrast, low-loss photonic platforms, such as the recently developed InGaP-on-insulator systems \cite{thiel2024wafer} ($n_\text{InGaP} = 3.4$), can enable stronger optical confinement with narrow optical linewidths. 
This could further lower the nonlinear threshold by at least an order of magnitude, paving the way for few- ($N<10$) or even single-polariton nonlinearity.
Such enhanced polariton nonlinearity would enable the development of integrated quantum polaritonic devices, such as single-photon sources and quantum logic gates for quantum information processing.
Furthermore, this platform could facilitate the large-scale integration of nonlinear polaritonic devices on photonic chips, with potential applications such as all-optical neural networks \cite{shen2017deep}, effectively addressing the persistent challenge of limited optical nonlinearity in conventional optical materials.

\section*{Author Contributions}
L.H. and B.Z. conceived the idea. Z.W. fabricated the device. Z.W. and L.H. performed the experiment. All authors discussed the results. L.H. and B.Z. wrote the paper with contributions from all authors. B.Z. supervised the project. 

\begin{acknowledgments}
We thank Adina Ripin and Mo Li for their assistance in preparing the 2D stack. This work was partly supported by the US Office of Naval Research (ONR) through grant N00014-20-1-2325 on Robust Photonic Materials with High-Order Topological Protection and grant N00014-21-1-2703, as well as the Sloan Foundation.
\end{acknowledgments}

\bibliography{apssamp}

\end{document}